\documentclass[prc,twocolumn,showpacs,preprintnumbers,amsmath,amssymb,superscriptaddress,floatfix,nofootinbib]{revtex4}
\usepackage{epsfig}
\usepackage{multirow}
\long\def\symbolfootnote[#1]#2{\begingroup%
\def\thefootnote{\fnsymbol{footnote}}\footnote[#1]{#2}\endgroup}
\newcommand{\smsm}[1]{\scriptscriptstyle{\scriptscriptstyle{#1}}}
\newcommand{\dslash}[1]{#1\!\!\!/}

  \def\CL{{\cal L}}
\def\CM{{\cal M}}

\def\ibid{{\it ibid.\ }}

\def\prl#1{Phys.\ Rev.\ Lett.\ {\bf #1}}

\def\prc#1{Phys.\ Rev.\ C\ {\bf #1}}
\def\prd#1{Phys.\ Rev.\ D\ {\bf #1}}

\def\plb#1{Phys.\ Lett.\ B\ {\bf #1}}

\def\etal{{\em et al.}}
\def\epja#1{Eur.\ Phys.\ J.\ A {\bf #1}}

\def\be{\begin{equation}}
\def\ee{\end{equation}}
\def\Be{\begin{eqnarray}}
\def\Ee{\end{eqnarray}}
\def\ba{\begin{array}}
\def\ea{\end{array}}

\begin{document}
\title{Studying $\Lambda^*$ resonances in the $p \bar p \rightarrow  \Lambda \bar\Lambda \eta$ reaction}

\author{Bo-Chao Liu} \email{liubc@xjtu.edu.cn} \author{Ke Wang}\affiliation{School of Physics, Xi'an
Jiaotong University, Xi'an, Shannxi 710049, China}

\begin{abstract}
In this work, we make a theoretical study on the $p \bar p
\rightarrow  \Lambda \bar\Lambda \eta$ reaction for antiproton beam
energy from threshold to 4GeV within an effective Lagrangian
approach and isobar model. By assuming this reaction is dominated by
the excitation of $\Lambda$ and $\bar \Lambda$ resonances in
intermediate states, we calculate the total cross sections and give
the predictions of the angular distribution and invariant mass
spectrum of final particles. In particular, we discuss the
possibility to verify the existence of a narrow $\Lambda$ resonance
found in the process of $K^- p\to \eta \Lambda$ in the present
reaction. It shows that the  $p \bar p \rightarrow \bar \Lambda
\Lambda \eta$ reaction can provide us with valuable information
about the $\Lambda$ resonances having significant couplings to $\bar
K N$ and $\Lambda\eta$ channels. Thus the experimental data of this
reaction will be a good supplement to the $\bar K N\to\eta \Lambda$
scattering data for studying $\Lambda$ resonances.

\end{abstract}
\maketitle

\section{Introduction}
The study on the properties of $\Lambda$ resonances constitutes one
important part of the research in the baryon spectroscopy, which
offers us useful information about the strong interaction in the
nonperturbative energy region and also tests of our knowledge in the
strange particle channels. Up to now, most of the knowledge about
$\Lambda$ resonances is from the analysis of the data in the $\bar K
N$ and $\pi\Sigma$ channels. Studies on other channels, although
very important, are still relatively lacking. Due to isospin
conservation, the $\eta\Lambda$ channel is of special interest
because it only couples to $\Lambda$ resonance, which offers a
relatively clean channel for studying the properties of the
$\Lambda$ resonances. But even with this advantage, the status of
current knowledge on the coupling of $\Lambda$ resonances to $\eta
\Lambda$ channel is still not satisfying. In the Particle Data
Group(PDG) book\cite{pdg}, there is only one $\Lambda^*$ state, i.e.
$\Lambda(1670)$, has well-established coupling with $\eta\Lambda$
channel. The decay branch ratio of other $\Lambda$ resonances to
this channel is still not well identified. It is possible that other
resonances indeed have weak coupling with this channel and are
therefore hard to study their couplings with $\eta\Lambda$. However,
the relatively poor quality of experimental data in this channel is
also a potentially important reason.

The Crystal Ball Collaboration data on the reaction $K^- p\to
\eta\Lambda$ near threshold published in 2012 have much higher
accuracy than before, which offers a good basis to investigate the
reaction mechanism of this reaction and to extract the properties of
$\Lambda$ resonances in the $\eta\Lambda$ channel. Based on the new
data an analysis within an effective Lagrangian approach and isobar
model was performed in Refs.\cite{liu1,liu2}. The main findings are,
although the $\Lambda(1670)$ gives the dominant contribution near
threshold, the bowl structure appearing in the angular distribution
may indicate a new narrow resonance. It was shown that the
experimental data supported the existence of a $D_{03}$ resonance
with M=$1668.5\pm 0.5$ MeV and $\Gamma=1.5\pm 0.5$ MeV(denoted as
$\Lambda^*_D$ for convenience). Due to the very narrow width, this
$\Lambda$ resonance is obviously not any existing $\Lambda$
resonance in the PDG book. The possible existence of a narrow
$\Lambda$ resonance in this channel was confirmed by another group
based on a coupled-channel analysis\cite{kamano1,kamano2}. However,
in their analysis the proposed narrow resonance has the quantum
numbers $J^P=\frac{3}{2}^+$(hereafter referred to as $\Lambda_P^*$).
Very interestingly, a narrow enhancement lying near the
$\eta\Lambda$ threshold was also found in the mass spectrum of
$K^-p$ in the decay of $\Lambda_c\to p K^- \pi^+$ at
Belle\cite{belle}. Until now, the origin of this enhancement is
still not well identified. Very recently, it was argued that the
enhancement might be caused by kinematical singularity\cite{xhliu}.
However, as stated by the authors of this work, partial wave
analysis is still needed to distinguish various scenarios.
Obviously, to establish whether the narrow resonance exists or not,
further studies on both theoretical and experimental sides are still
needed.

The $\rm \bar PANDA$ experiment\cite{panda} at the Facility for
Antiproton and Ion Research (FAIR) will be carried out in the near
future, which is well suited for exploring the spectroscopy of
strange and charmed baryons. Such experiment will definitely offer
valuable data for improving our knowledge of the strong interaction
and of hadron spectroscopy. Encouraged by the prospect, there have
been a series of theoretical investigations on the new opportunities
for studying the baryon spectroscopy in $N\bar N$
collisions\cite{he,haidenbauer,shyam,dmli,haidenbauer2,liuxiang}. In
their studies, they mainly focused on the production of charmed
hadrons. In this work, we attempt to show that the reaction $p \bar
p \rightarrow  \Lambda\bar \Lambda \eta$ may be a suitable place to
explore the properties of $\Lambda$ resonances. To our best
knowledge, there is still no experimental data available for this
reaction. Our calculation will mainly be based on an effective
Lagrangian approach and isobar model. In our model, the $\Lambda$
resonances are excited due to the $K$ and $K^*$ meson exchanges
between the initial proton and antiproton. Thus this reaction offers
the possibility to explore the $\Lambda$ resonances having
significant coupling with $\bar KN$($\bar K^*N$) and $\Lambda\eta$
channels. Till now, only the $\Lambda(1670)$ is known to have
significant coupling to these channels. If the narrow resonance
mentioned above indeed exists, it should also play a role in this
reaction. In addition, the present reaction may also proceed through
the excitation of mesonic resonances in the intermediate states,
which finally decay to $\Lambda\bar\Lambda$. In our model, the
production of such states is induced by the exchange of nucleon or
nucleon resonance(e.g. $N(1535)$) between initial proton and
antiproton. For nucleon exchange, its contribution should be
suppressed due to the vanishing $NN\eta$ coupling\cite{na3}. While
for nucleon resonance exchange, its contribution can not be well
estimated due to the poor knowledge of the $\bar N N^*\to
\Lambda\bar\Lambda$ process. Thus we choose to ignore these
contributions\footnote{For the purpose of studying the $\Lambda$
resonances, the uncertainties due to these contributions can be
controlled experimentally. When some resonance decaying to the
$\Lambda\bar\Lambda$ channel contributes significantly and its
contribution overlap with the narrow $\Lambda$ resonance, it is
possible to separate their contributions or their bands in the
Dalitz plot by choosing a different beam energy.}. In this work, we
shall consider the contributions from the $\Lambda(1670)$ and the
possible narrow $\Lambda$ resonance with considering both the
$P_{03}$ and $D_{03}$ assignments for its quantum numbers. Most of
the model parameters are determined by fitting the data of the $K^-
p\to \eta\Lambda$ reaction. The predictions of the angular
distribution, invariant mass spectrum and Dalitz plot are presented,
which should be useful for future comparisons with data and looking
for the possible narrow $\Lambda$ resonance.

This paper is organized as follows. In Sec. II, the theoretical
framework and amplitudes are presented for the reaction $p \bar p
\rightarrow \bar \Lambda \Lambda \eta$. In Sec. III, the numerical
results are presented with some discussions. Finally, the paper ends
with a short summary in Sec. IV.

\section{Theoretical Formalism}
\begin{figure}[htbp]
\begin{center}
\includegraphics[scale=0.6]{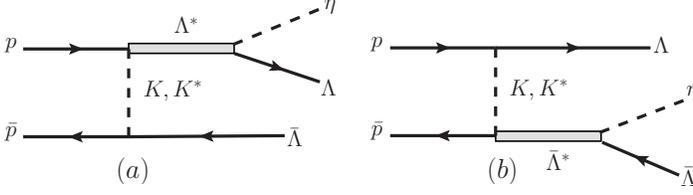}
\caption{Model for the reaction $p \bar p \to \Lambda \bar\Lambda \eta$.} \label{feyn_fig}
\end{center}
\end{figure}

In this work, we investigate the $p \bar p \rightarrow  \Lambda
\bar\Lambda \eta$ reaction  within an effective Lagrangian approach
and isobar model. We assume that this reaction is dominated by the
excitation of $\Lambda$ and $\bar\Lambda$ resonances in the
intermediate states with considering the contributions from the
$\Lambda(1670)$ and a very narrow $\Lambda^*_D$/$\Lambda^*_P$
resonance suggested in Refs.\cite{liu1,liu2,kamano1,kamano2}. The
basic Feynman diagrams are depicted in Fig.\ref{feyn_fig}.

The effective Lagrangians describing the $KN\Lambda$ and
$K^*N\Lambda$ interactions can be given as \Be\label{ksnl}
\CL_{KN\Lambda}&=&-ig_{KN\Lambda}\bar \Psi_N\gamma_5
\Psi_{\Lambda}\Phi_K +h.c. \Ee \Be\label{kssnl}
\CL_{K^*N\Lambda}=g_{\smsm{K^*N\Lambda}}\bar \Psi_N(\gamma^\mu
K^*_\mu -
\frac{\kappa_{\smsm{K^*}}}{2m_{\smsm{N}}}\sigma^{\mu\nu}\partial_\nu
K^*_{\mu}) \Psi_{\Lambda} +h.c.\Ee
 The value of $g_{NK\Lambda}$ can be determined by the SU(3) predictions, and we adopt $g_{NK\Lambda}=-13.24$ in our
calculations\cite{oh1,oh2}. For the coupling constants
$g_{K^*N\Lambda}$ and $\kappa_{\smsm{K^*}}$, we take their values
from the Nijmegen potential, i.e. $g_{K^*N\Lambda}=4.26$ and
$\kappa_{\smsm{K^*}}=2.66$\cite{nijmegen1,nijmegen2}.

The relevant interaction Lagrangians involving the $\Lambda(1670)$
or the $\Lambda^*_D$/$\Lambda^*_P$ resonances are used in the same
forms as in Refs.\cite{liu1,liu2},
 \Be
\CL_{\smsm{\Lambda(1670)\Lambda\eta}}&=&g_{\smsm{\Lambda(1670)\Lambda\eta}}\bar \Lambda\Lambda^*\eta +H.c.,\\
{\cal L}_{\smsm{\Lambda(1670) KN}}&=&g_{\smsm{\Lambda(1670) KN}} \bar\Lambda^* \bar{K} N +H.c., \\
{\cal L}_{\smsm{\Lambda(1670) K^*N}}&=&ig_{\smsm{\Lambda(1670) K^*N}} \bar\Lambda^*\gamma_5\gamma^\mu K^*_\mu N +H.c., \\
{\cal L}_{\smsm{\Lambda^*_DKN}} &=& {f_{\smsm{\Lambda^*_DKN}}\over
m_K}\partial_\mu\bar{K}\bar\Lambda^{*\mu} \gamma_5
N+H.c.,\\
{\cal L}_{\smsm{\Lambda^*_D\Lambda\eta}} &=& {
f_{\smsm{\Lambda^*_D\Lambda\eta}}\over
m_\eta}\partial_\mu\eta{\bar\Lambda}^{*\mu}\gamma_5{\Lambda}+H.c.,\\
{\cal L}_{\smsm{\Lambda^*_PKN}} &=& {f_{\smsm{\Lambda^*_PKN}}\over
m_K}\partial_\mu\bar{K}\bar\Lambda^{*\mu}
N+H.c.,\\
{\cal L}_{\smsm{\Lambda^*_P\Lambda\eta}} &=& {
f_{\smsm{\Lambda^*_P\Lambda\eta}}\over
m_\eta}\partial_\mu\eta{\bar\Lambda}^{*\mu}{\Lambda}+H.c..
 \Ee
The coupling constant $g_{\smsm{\Lambda(1670) K^*N}}=0.753$ is taken
from Ref.\cite{xiao}, where its value is obtained based on a chiral
quark model. For the $\Lambda_D^*\bar KN$ and
$\Lambda_D^*\eta\Lambda$ couplings, we follow the results in
Refs.\cite{liu1,liu2}, where the relevant coupling constants are
fitted to the experimental data of the $K^-p\to\eta\Lambda$
reaction(Scenario I). The obtained parameters are shown in Table
\ref{tab1}. To get the parameters for the $P_{03}$ assignment, we
fit them to the same data set as in the Scenario I but assuming the
new $\Lambda$ resonance is a $P_{03}$ state(Scenario II). The
obtained mass and width of the narrow resonance are consistent with
the results in Ref.\cite{kamano2} within uncertainties. Note that we
also calculate the predictions of the $\Lambda$ polarization for the
$K^-p\to \eta\Lambda$ reaction and find that in Scenario II the
predictions seem incompatible to the available data(see also Fig. 20
of Ref.\cite{kamano1}). So more accurate $\Lambda$ polarization data
of the $K^-p\to\eta\Lambda$ reaction will be helpful to clarify the
quantum numbers of this narrow resonance.

Because hadrons cannot be treated as elementary particles in the
energy region under study, it is necessary to take into account the
internal structures and off-shell effects. In phenomenological
models, this is usually done by introducing form factors. In this
work, we adopt the following form factor for various meson exchange
vertices, \Be F_M(q)&=&\frac{\Lambda^2_M- m^2}{\Lambda^2_M-q^2}, \Ee
where $\Lambda_M$, $m$ and $q$ are the cutoff parameter, the mass of
the exchanged particle and the exchanged momentum. The cutoff
parameters for the $KN\Lambda$ and $K^*N\Lambda$ vertices are
adopted as $\Lambda_K=1.1$ GeV and $\Lambda_{K^*}=0.9$
GeV\cite{oh1,oh2}, respectively. While, the cutoff parameters for
the $\Lambda^*\bar KN$ vertices are not well determined in
literatures. In present work, we use the same value as that for the
$\Lambda\bar KN$ vertex and the uncertainties due to this parameter
will be discussed in the next section.

The propagators for the $\Lambda(1670)$, $\Lambda^*_{P/D}$, $K$ and
$K^*$ are adopted as the following forms: \Be
G_{\smsm{\Lambda(1670)}}^{\pm}(q)&=&\frac{i(\dslash{q}
\pm M_{\Lambda(1670)})}{q^2-M^2_{\Lambda(1670)}+iM_{\Lambda(1670)}\Gamma_{\Lambda(1670)}}\,,\\
G_{\Lambda^*_{P/D}}^{\pm}(q)&=&\frac{i(\dslash{q}
\pm M_{\Lambda^*_{P/D}})}{q^2-M^2_{\Lambda^*_{P/D}}+iM_{\Lambda^*_{P/D}}\Gamma_{\Lambda^*_{P/D}}}[-g_{\mu\nu}+\frac{1}{3}\gamma_\mu\gamma_\nu \nonumber\\
 &&\pm \frac{1}{3M_{\Lambda^*_{P/D}}}(\gamma_\mu q_\nu-\gamma_\nu q_\mu ) +\frac{2}{3M_{\Lambda^*_{P/D}}^2}q_\mu q_\nu ]\,,\\
G_{K}(q)&=& \frac{i}{q^2-m^2_{K}}, \\
G_{K^*}^{\mu\nu}(q)&=&i\frac{-g^{\mu\nu}+q^\mu
q^\nu/m^2_{K^*}}{q^2-m^2_{K^*}} \Ee where the superscript + and -
correspond to particle and antiparticle respectively.

With the ingredients given above, the amplitudes for various
diagrams can be written by following the Feynman rules. Here we
present the individual amplitudes explicitly, \Be
\CM_{a,K}^{\smsm{\Lambda(1670)}}&=g_{\smsm{\Lambda(1670)\Lambda\eta}}g_{\smsm{NK\Lambda}}g_{\smsm{\Lambda^*
{\bar K}N}} \bar u_{\Lambda ,
s_\Lambda}G_{\Lambda^*}^{(+)}(P)F_{K}(q)\nonumber\\& u_{p ,s_p}
 G_K(q) \bar v_{\bar p, s_{\bar p}}\gamma_5 v_{\bar \Lambda ,s_{\bar \Lambda}} \nonumber \\
\CM_{b,K}^{\smsm{\Lambda(1670)}}&=-g_{\smsm{\Lambda(1670)\Lambda\eta}}g_{\smsm{NK\Lambda}}g_{\smsm{\Lambda^*
{\bar K}N}} \bar v_{\bar p , s_{\bar p}}G_{\Lambda^*}^{(-)}(P')F_{K}(q')\nonumber\\& v_{\bar \Lambda, s_{\bar \Lambda}} G_K(q') \bar u_{\Lambda, s_{\Lambda}}\gamma_5 u_{p ,s_{p}} \nonumber \\
\CM_{a,K^*}^{\smsm{\Lambda(1670)}}&=-g_{\smsm{\Lambda(1670)\Lambda\eta}}g_{\smsm{K^*N\Lambda}}g_{\smsm{\Lambda^*
K^*N}} \bar u_{\Lambda ,
s_\Lambda}G_{\Lambda^*}^{(+)}(P)\gamma_5\gamma_\mu F_{K^*}(q)\nonumber\\
& u_{p ,s_p}
 G^{\mu\nu}_{K^*}(q)\nonumber  \bar v_{\bar p, s_{\bar p}}(\gamma_\nu - i\frac{\kappa_{\smsm{K^*}}}{2m_N}\sigma_{\nu\rho}q^\rho ) v_{\bar \Lambda ,s_{\bar \Lambda}} \nonumber \\
\CM_{b,K^*}^{\smsm{\Lambda(1670)}}&=g_{\smsm{\Lambda(1670)\Lambda\eta}}g_{\smsm{K^*N\Lambda}}g_{\smsm{\Lambda^*
K^*N}} \bar v_{\bar p , s_{\bar p}}\gamma_5\gamma_\mu
F_{K^*}(q')G_{\Lambda^*}^{(-)}(P') \nonumber
\\&  v_{\bar \Lambda, s_{\bar \Lambda}} G_{K^*}^{\mu\nu}(q') \bar
u_{\Lambda, s_{\Lambda}}
(\gamma_\nu - i\frac{\kappa_{\smsm{K^*}}}{2m_N}\sigma_{\nu\rho} q'^\rho ) u_{p ,s_{p}} \nonumber \\
\CM_{a}^{\Lambda^*_D}&=e^{i\phi_\alpha}{g_{NK\Lambda}f_{\Lambda^*_D\bar
KN} f_{\Lambda^*_D\Lambda\eta}\over m_\eta m_K} \bar u_{\Lambda ,
s_\Lambda}\gamma_5G_{\Lambda^*}^{(+)\mu\nu}(P)p_{\mu}^\eta q_{\nu}\nonumber \\
& F_{\Lambda^*}(q)\gamma_5 u_{p ,s_{p}}  G_K(q) \bar v_{\bar p, s_{\bar p}}\gamma_5 v_{\bar \Lambda ,s_{\bar \Lambda}} \nonumber \\
\CM_{b}^{\Lambda^*_D}&=-e^{i\phi_\alpha}{g_{NK\Lambda}f_{\Lambda^*_D\bar
KN} f_{\Lambda^*_D\Lambda\eta}\over m_\eta m_K} \bar v_{\bar p ,
s_{\bar p}}\gamma_5
G_{\Lambda^*}^{(-)\mu\nu}(P')p_{\nu}^\eta\nonumber \\ &
q'_{\mu}F_{\Lambda^*}(q')\gamma_5 v_{\bar \Lambda ,s_{\bar \Lambda}}
  G_K(q') \bar u_{\Lambda,
s_{\Lambda}} \gamma_5 u_{p ,s_{p}}\nonumber\\
\CM_{a}^{\Lambda^*_P}&=e^{i\phi_\alpha}{g_{NK\Lambda}f_{\Lambda^*_P\bar
KN} f_{\Lambda^*_P\Lambda\eta}\over m_\eta m_K} \bar u_{\Lambda ,
s_\Lambda}G_{\Lambda^*}^{(+)\mu\nu}(P)p_{\mu}^\eta q_{\nu} \nonumber \\
&F_{\Lambda^*}(q) u_{p ,s_{p}}  G_K(q) \bar v_{\bar p, s_{\bar p}}\gamma_5 v_{\bar \Lambda ,s_{\bar \Lambda}} \nonumber \\
\CM_{b}^{\Lambda^*_P}&=-e^{i\phi_\alpha}{g_{NK\Lambda}f_{\Lambda^*_P\bar
KN} f_{\Lambda^*_P\Lambda\eta}\over m_\eta m_K} \bar v_{\bar p ,
s_{\bar p}} G_{\Lambda^*}^{(-)\mu\nu}(P')p_{\nu}^\eta
q'_{\mu}\nonumber \\ &F_{\Lambda^*}(q') v_{\bar \Lambda ,s_{\bar
\Lambda}}
  G_K(q') \bar u_{\Lambda,
s_{\Lambda}} \gamma_5 u_{p ,s_{p}} \nonumber\Ee In the above
formulas, the letters in the parentheses indicate the momentum of
the exchanged particles and $p_\eta$ denotes the momentum of the
$\eta$ in the final state.

Based on the individual scattering amplitudes presented above, the
general differential cross section of $p\bar p\to
\Lambda\bar\Lambda\eta$ reads
\begin{eqnarray}
d\sigma&=&\frac{1}{16}\frac{1}{\sqrt{(p_p\cdot p_{\bar
p})^2-m_p^4}}\frac{1}{(2 \pi)^5} \sum_{s_i,s_f} |{\cal
M}_{fi}|^2\nonumber \\&&\cdot\prod^3_{a=1}\frac{ d^{3}
p_{a}}{2E_{a}} \delta^4 (P_i - P_f), \label{eqcs}
\end{eqnarray}
where $\CM_{fi}$ represents the total amplitude, $P_i$ and $P_f$
represent the sum of all the momenta in the initial and final
states, respectively. $p_a$ denotes the momenta of the three
particles in the final state.

\begin{table*}
\caption{Parameters obtained by fitting to the total and
differential cross sections of the $K^-p\to \eta\Lambda$ reaction.}
\begin{tabular}{|c|c|c|c|c|c|c|}
\hline
\hline Scenario &Considered resonance   & Product of coupling constants &Relative phase($\phi_\alpha$)&Mass(MeV) &Width(MeV)&$\chi^2/dof$\\
 \hline
\multirow{2}*{I}& $\Lambda(1670)$& $g_{\Lambda(1670)\Lambda\eta}
g_{\Lambda(1670) {\bar K}N}=0.30\pm 0.03$& $0.$& $1672.5\pm 1.0$&
$24.5\pm 2.7$&\multirow{2}*{0.88}
\\ \cline{2-6}
&$\Lambda_D^*$& $f_{\Lambda_D^*\bar KN}
f_{\Lambda_D^*\Lambda\eta}=28.2\pm 2.4$& $5.66\pm 0.47$& $1668.5\pm
0.5$& $1.5\pm 0.5$ &
\\ \hline
\multirow{2}*{II}& $\Lambda(1670)$& $g_{\Lambda(1670)\Lambda\eta}
g_{\Lambda(1670) {\bar K}N}=0.32\pm 0.03$& $0.$& $1672.2\pm 0.8$&
$27.6\pm 1.0$ &\multirow{2}*{0.86}\\ \cline{2-6} &$\Lambda_P^*$&
$f_{\Lambda_P^*\bar KN} f_{\Lambda_P^*\Lambda\eta}=2.98\pm 0.08$&
$0.61\pm 0.08$& $1663.6\pm 0.5$& $11.0\pm 1.4$& \\ \hline III
&$\Lambda(1670)$& $g_{\Lambda(1670)\Lambda\eta} g_{\Lambda(1670)
{\bar K}N}=0.28\pm 0.02$& $-$& $1671.5\pm 0.2$& $23.2\pm 0.2$& 1.22
\\ \hline\hline
\end{tabular}
\label{tab1}
\end{table*}

Before presenting the calculated results, we need to discuss the
possible effects of the $p\bar p$ initial state interaction(ISI) and
$\Lambda \bar\Lambda$ final state interaction(FSI) in the present
reaction. It is known that the ISI may have important effects on the
meson production in nucleon-nucleon collisions\cite{fsi1,fsi2},
where the ISI reduces the cross section by an over factor with
slight energy dependence. The study on the $p\bar p\to
\Lambda_c\bar\Lambda_c$ reaction also shows that the ISI effect may
reduce the cross section by a factor of 100\cite{haidenbauer}. So it
is natural to expect that the ISI effect may also be important for
the reaction under study. When we consider the energy region near
threshold, the interaction between final $\Lambda$ and $\bar
\Lambda$ may also become important\cite{watson}. A reliable
description of the FSI between the $\Lambda$ and $\bar\Lambda$ will
rely on a good understanding of the $\Lambda\bar\Lambda$
interaction, for which our knowledge is still rather limited due to
the absence of data. Therefore an accurate description of the FSI
between $\Lambda$ and $\bar\Lambda$ is still not possible. To take
into account the ISI effect, we adopt a phenomenological approach as
in Refs.\cite{isi1,isi2}. Interestingly, in a recent
work\cite{shyam}, the authors have adopted the same approach and
applied it to study the $p\bar p\to\bar\Lambda_c^-\Lambda_c^+$
reaction. In their work, the parameters for the ISI were checked by
reproducing the near threshold cross sections predicted by Juelich
model, in which model ISI is taken into account more rigorously.
Using the same parameters, they can also successfully reproduce the
cross sections of the $p\bar p\to \Lambda\bar\Lambda$ reaction near
threshold without considering FSI effect explicitly. For simplicity,
in this work we choose to follow the approach in Ref.\cite{shyam}
and adopt their parameters. Thus we assume the effect of FSI has
been effectively absorbed into the model parameters. Here we want to
note even though we treat the ISI and FSI in a model dependent way,
the main conclusions of the present work should not be changed
significantly since our primary goal is to have an order of
magnitude estimation of the total cross sections and to investigate
the relative importance of various $\Lambda$ resonances in this
reaction.
\section{RESULTS AND DISCUSSIONS}


With the formulas and ingredients given in last section, the total
and differential cross sections can be calculated in a
straightforward way and we present the results in this section. To
investigate the roles of the $\Lambda(1670)$ and the possible narrow
resonance in the $p\bar p\to \Lambda\bar\Lambda\eta$ reaction, we
will consider three scenarios. First, we include the contributions
from both the $\Lambda(1670)$ and the narrow $\Lambda^*_D$ in the
reaction(Scenario I). Second, we adopt the assumption that the
narrow resonance is a $P_{03}$ state as in
Refs.\cite{kamano1,kamano2} and consider its contribution in this
reaction(Scenario II). Finally, we consider the case that the narrow
resonance does not exist and thus there is no contribution from the
narrow resonance(Scenario III). For all the three scenarios, the
parameters of the models such as the coupling constants and relative
phases are determined by fitting the total and differential cross
sections of the $K^-p\to\eta\Lambda$ reaction, where these
resonances play important roles. The adopted parameters have been
listed in Table \ref{tab1}.

\begin{figure}[htbp]
\begin{center}
\includegraphics[scale=0.28]{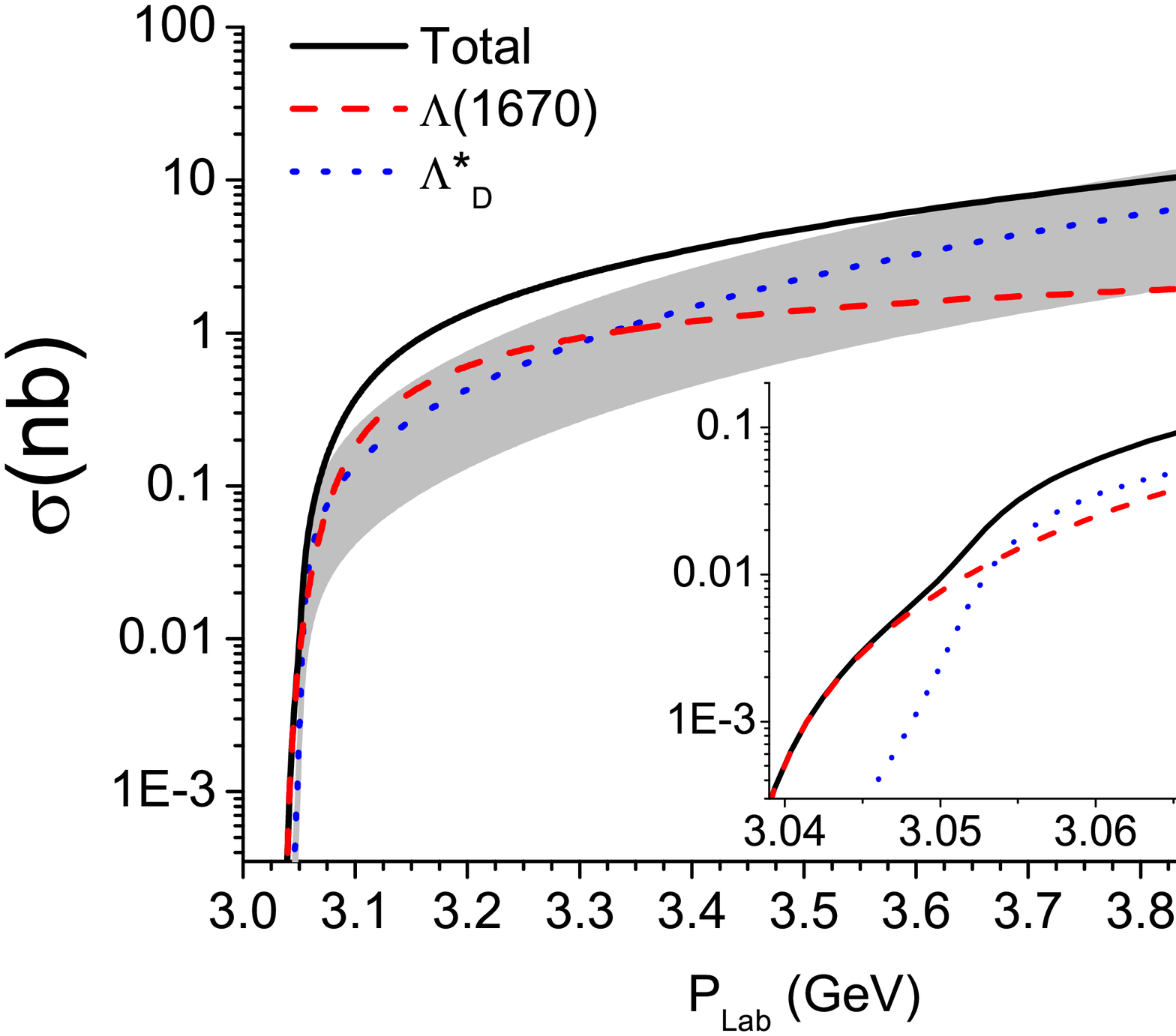}
\includegraphics[scale=0.28]{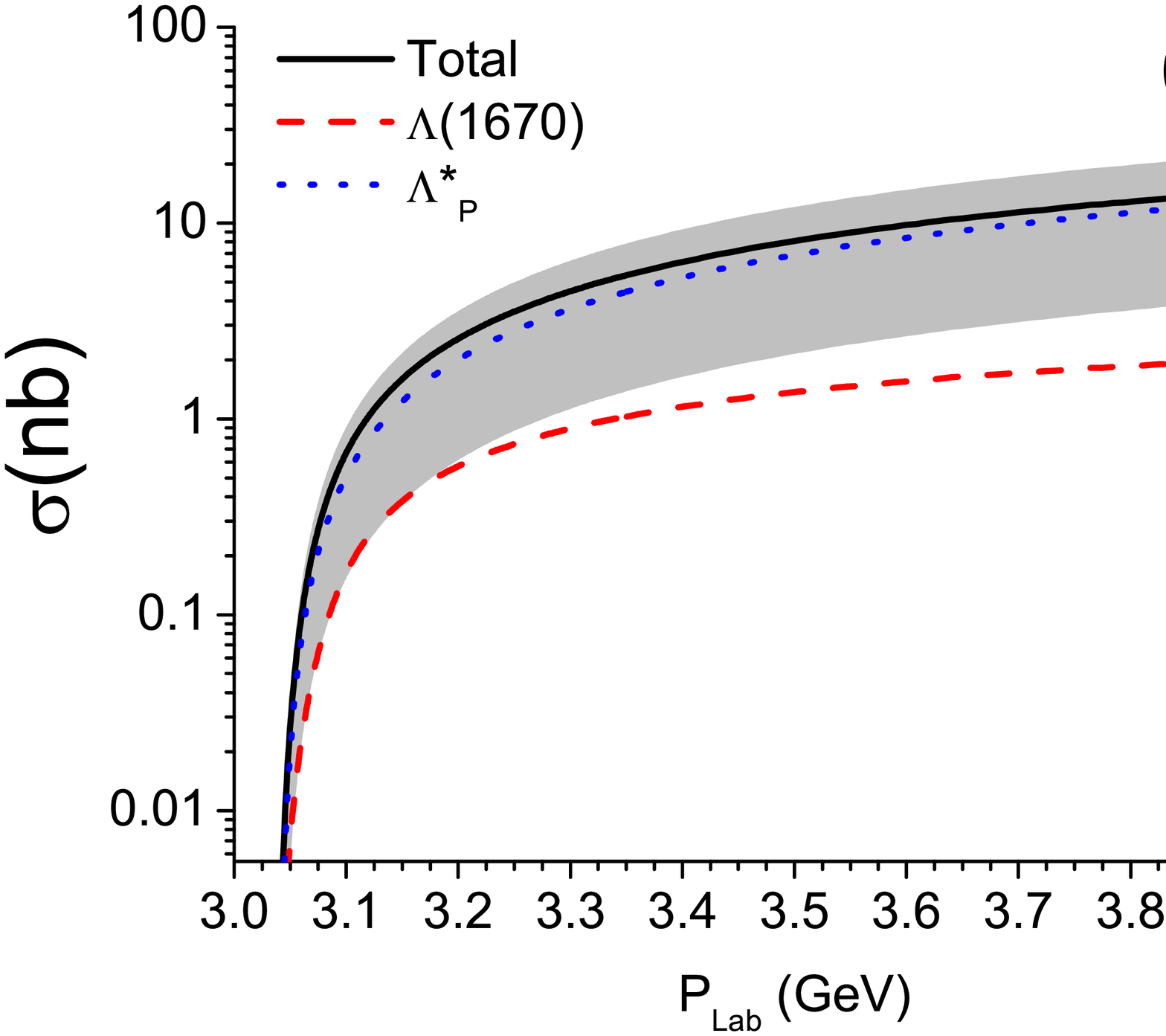}
\caption{Total cross sections as a function of the antiproton beam
momentum obtained in Scenario I (a) and Scenario II (b). The band
corresponds to the results of $\Lambda_{P/D}^*$ by varying the
cutoff parameter for the $\Lambda^*_{P/D}\bar KN$ vertex from 0.8 to
1.4 GeV.} \label{xsection}
\end{center}
\end{figure}

In Fig.\ref{xsection}a, we show the total cross sections and the
individual contributions of various resonances in Scenario I and II.
As shown in Fig.\ref{xsection}a, the $\Lambda_D^*$ contribution is
suppressed at the very near threshold region and the $\Lambda(1670)$
gives the most important contribution at first. When the center of
mass energy approaches the threshold of $\Lambda_D^*\bar
\Lambda$/$\bar\Lambda_D^* \Lambda$
($\sqrt{s}=M_{\Lambda_D^*}+M_{\bar\Lambda}$), the production of
$\Lambda_D^*$ starts to play more important role and causes the
strong energy dependence at around $P_{Lab}=3.05$ GeV. As the energy
increases and comes near the
$\Lambda(1670)\bar\Lambda/\bar\Lambda(1670)\Lambda$ threshold, the
contribution of the $\Lambda(1670)$ becomes dominant again. But at
higher energies, the $\Lambda^*_{D}$'s contribution exceeds the
contribution of the $\Lambda(1670)$ once again. It should be noted
that at the near threshold region the energy dependence due to FSI
should also play an important role. Since such effects are not
considered in this work, the discussions presented above only serve
to show the production of $\Lambda_D^*$ may cause significant energy
dependence in the total cross sections. Compared to the role of the
$\Lambda_D^*$ in the $K^-p\to\eta\Lambda$ reaction, its role is
significantly enhanced in the present reaction. The enhancement is
mainly due to the D-wave nature of the $\Lambda^*_D\bar KN$ coupling
and the large threshold momentum of the present reaction. At the
near threshold region, the vertex function of the $\Lambda^*_D\bar
KN$ vertex is roughly proportional to $p_{th}^2$, where $p_{th}$ is
the threshold momentum of the reaction in the center of mass frame.
Therefore, the large threshold momentum of the present reaction
makes the contribution of the $\Lambda^*_D$ much more significant
than that in the $K^-p\to\eta\Lambda$ reaction, where the threshold
momentum is a factor of 4 smaller. There is no such enhancement for
the s-wave state $\Lambda(1670)$, so the role of the $\Lambda^*_D$
becomes more important in the present reaction. The final results
for the $\Lambda^*_D$ contribution certainly still rely on other
model parameters. In our model, it may come from the cutoff
parameter for the $\Lambda^*\bar KN$ vertex. To obtain the results
shown in Fig.\ref{xsection}a, we have adopted $\Lambda_K=1.1$ GeV in
the $\Lambda^*_D\bar KN$ vertex as that for the $\Lambda(1670)\bar
KN$ vertex in the calculations. To check the dependence on this
parameter, we have also allowed the cutoff parameter for the
$\Lambda^*_D\bar KN$ vertex varying from $0.8$ to $1.4$ GeV, which
results in the band shown in the figure. In Scenario II, our
results(Fig.\ref{xsection}b) show that the production of the
$\Lambda_P^*$ dominates this reaction even at the near threshold
region. In fact, in our fitting of the $K^-p\to\eta\Lambda$ reaction
data, we also find, even though the $\Lambda(1670)$ gives the
dominant contribution, the $\Lambda_P^*$ contribution is significant
as well. Compared to the s-wave state $\Lambda(1670)$, the
contribution of the $\Lambda^*_P$ in the present reaction is also
enhanced due to the large threshold momentum as mentioned above. The
band in the Fig.\ref{xsection}b shows the uncertainties due to the
cutoff parameter for the $\Lambda^*_P\bar KN$ vertex by varying it
from 0.8 to 1.4 GeV.

\begin{figure}[htbp]
\begin{center}
\includegraphics[scale=0.5]{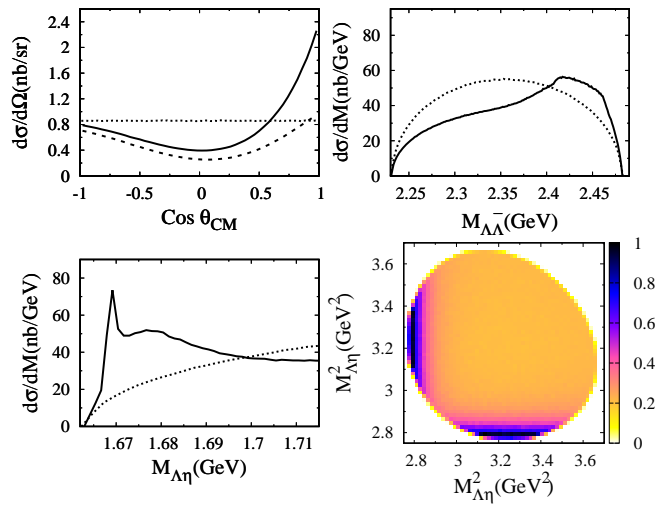}
\caption{Predictions for the angular distribution of $\eta$ in the
$\Lambda\eta$ rest frame, the spectrum of $M_{\Lambda\bar \Lambda}$,
the spectrum of $M_{\Lambda\eta}$ and Dalitz plot for Scenario I at
$P_{lab}=3.84$ GeV. The predicted results are shown by the solid
lines and compared with the phase space distribution(dotted lines).
The dashed line of the $\eta$ angular distribution represents the
results with imposing a cut $M_{\bar\Lambda\eta}>1.75$ GeV on the
invariant mass of $\bar\Lambda\eta$.} \label{dwave}
\end{center}
\end{figure}

In Fig.\ref{dwave}, we show the differential cross sections obtained
in Scenario I at $P_{lab}=3.84$ GeV, where the FSI of
$\Lambda\bar\Lambda$ is expected to be small. As can be seen from
the figure, there is a sharp peak appearing in the $M_{\Lambda\eta}$
spectrum. Compared to the small bump shown in the total cross
sections of the $K^-p\to \eta\Lambda$ reaction(see Fig.2 of
Ref.\cite{liu2}), the signal of the $\Lambda_D^*$ is significantly
enhanced here as expected from the total cross sections shown in
Fig.\ref{xsection}a. The angular distribution of $\eta$ is studied
in the $\Lambda\eta$ rest frame and the $\theta_{CM}$ is defined as
the angle of the $\eta$ momentum relative to the beam direction. The
angular distribution shows an asymmetry at forward and backward
angles. The forward peak is mainly caused by the $\eta$ meson
originated from the decay of the $\bar \Lambda$ resonances. While,
the backward enhancement is caused by the $\Lambda_D^*$ and its
interference with other contributions. If we eliminate the
contribution from the $\bar\Lambda$ resonances by a cut with
requiring $M_{\bar\Lambda\eta}>1.75$GeV, the asymmetry can be
significantly reduced. The remaining concave-up shape of the angular
distribution indicates the higher partial wave contributions from
the $\Lambda_D^*$(see also Fig.\ref{swave} for comparison). We have
also checked even if we adopt $\Lambda_K=0.8$ GeV in the
calculations, there is still a clear bump relative to the
enhancement caused by the $\Lambda(1670)$ in the $M_{\Lambda\eta}$
spectrum.

\begin{figure}[htbp]
\begin{center}
\includegraphics[scale=0.5]{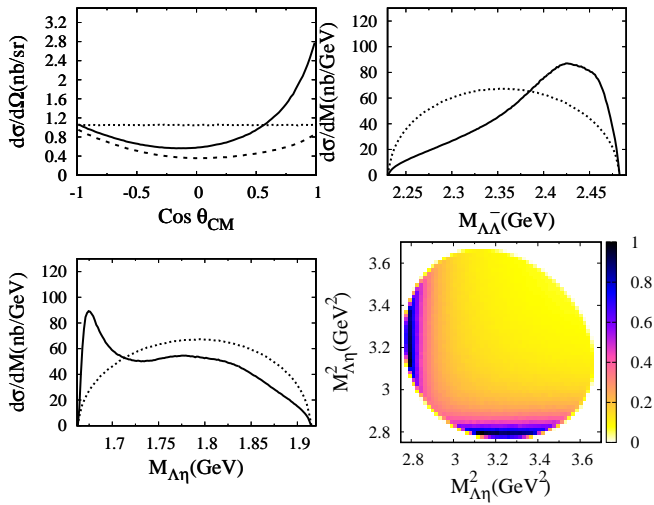}
\caption{Same as Fig.\ref{dwave} but for Scenario II.} \label{pwave}
\end{center}
\end{figure}

The corresponding results for Scenario II are presented in
Fig.\ref{pwave}. In this case, there is no clear peak of the
$\Lambda^*_P$ in the $M_{\Lambda\eta}$ spectrum. This is mainly
because the $\Lambda_P^*$ in our model lies very close to the
$\Lambda\eta$ threshold and has a relatively large width($\sim 11$
MeV). The enhancement in the $M_{\Lambda\eta}$ spectrum compared to
the phase space distribution is caused by a coherent sum of the
contributions of the $\Lambda(1670)$ and the $\Lambda^*_P$. On the
other hand, the angular distribution of $\eta$ shows distinct
features compared to the results without the $\Lambda_P^*$
contribution(Fig.\ref{swave}). After eliminating the $\bar\Lambda^*$
contribution as done for Scenario I, the structure shown in the
$\eta$ angular distribution in the $\Lambda\eta$ rest frame also
clearly indicates the higher partial wave contribution from the
$\Lambda_P^*$. Compared to the results shown in Fig.\ref{dwave}, we
find that the $\eta$ angular distributions have similar patterns in
these two cases. Thus it is difficult to identify the quantum
numbers of the narrow resonance by only analyzing the angular
distributions, and the polarization data may be needed. A detailed
study on the polarization observables will rely on a more rigorous
treatment of both ISI and FSI and is out of the scope of present
work. However, as can be seen from the figures an accurate
measurement of the Dalitz plot or invariant mass spectrums can still
offer valuable information about the narrow resonance, since the
Scenarios I and II predict distinct features in these observables.

\begin{figure}[htbp]
\begin{center}
\includegraphics[scale=0.5]{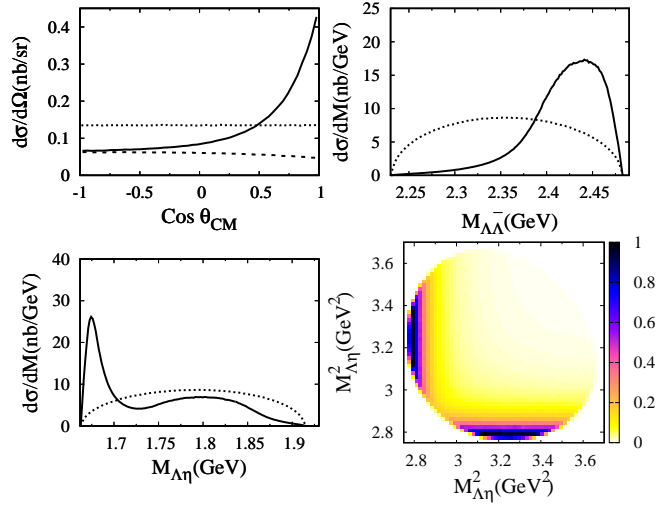}
\caption{Same as Fig.\ref{dwave} but for Scenario III.}
\label{swave}
\end{center}
\end{figure}
Finally, in Fig.\ref{swave} we show the results with only
considering the $\Lambda(1670)$ contribution at $P_{lab}=3.84$ GeV
for comparisons. As can be seen from the figure, there is a clear
enhancement caused by the $\Lambda(1670)$ appearing in the
$M_{\eta\Lambda}$ spectrum. The corresponding enhancement in the
Dalitz plot is also significant. After eliminating the
$\bar\Lambda^*$ contributions, the angular distribution of the final
$\eta$ in the $\eta\Lambda$ rest frame is roughly flat. Compared to
the corresponding $\eta$ angular distributions(dashed line) in
Scenario I and II, the significant curvature shown in the $\eta$
angular distribution in the rest frame of the $\eta\Lambda$ system
can be looked as the evidence for the existence of the new $\Lambda$
resonance. It is also worth noting that even though the $K^*$
exchange contribution is included in the calculations, we find its
contribution is very small and can be neglected.

\section{conclusion}
In this work, we investigate the production of $\Lambda/\bar\Lambda$
resonances in the $p\bar p\to \Lambda\bar\Lambda\eta$ reaction
within an effective Lagrangian approach and isobar model.
Especially, we investigate the possibility to verify the existence
of a new narrow $\Lambda$ resonance found in the
$K^-p\to\eta\Lambda$ reaction near threshold. Based on our model
calculations, we find the narrow resonance, if exists, can give
significant contribution in this reaction and the total cross
sections of this reaction is found to be roughly at the order of
$0.1\sim 10$ nb at $P_{lab}=3.1-4$ GeV. Thus the measurements of
this reaction will offer a good opportunity to verify the existence
of this resonance. The predictions can be tested in the future by
the $\rm\bar PANDA$ experiment.
\begin{acknowledgements} We acknowledge the support from the
National Natural Science Foundation of China under Grants No.
U1832160 and No. 11375137, the Natural Science Foundation of Shaanxi
Province under Grant No. 2015JQ1003 and No. 2019JM-025, and the
Fundamental Research Funds for the Central Universities.

\end{acknowledgements}

\end{document}